\documentclass[conference]{IEEEtran}
\IEEEoverridecommandlockouts

\usepackage{tikz}
\usetikzlibrary{shapes.geometric, arrows.meta, positioning, calc}
\usepackage{booktabs}
\usepackage[table]{xcolor}
\usepackage{hyperref}
\usepackage{cite}
\usepackage{amsmath,amssymb,amsfonts}
\usepackage{algorithmic}
\usepackage{graphicx}
\usepackage{textcomp}
\usepackage{xcolor}
\usepackage{listings}
\usepackage{url}
\def\BibTeX{{\rm B\kern-.05em{\sc i\kern-.025em b}\kern-.08em
    T\kern-.1667em\lower.7ex\hbox{E}\kern-.125emX}}
\usepackage{xcolor}

\begin{document}

\title{SCALAR: A Neurosymbolic Framework for Automated Conjecture and Reasoning in
Quantum Circuit Analysis}


\author{
\IEEEauthorblockN{Sean Feeney\textsuperscript{1*}, 
Pooja Rao\textsuperscript{2}, 
Andreas Klappenecker\textsuperscript{1}, 
Reuben Tate\textsuperscript{3}, 
Yuri Alexeev\textsuperscript{2}, 
Stefano Mensa\textsuperscript{2},
Elica Kyoseva\textsuperscript{2},
\\
and Stephan Eidenbenz\textsuperscript{3}}
\\[4pt]
\IEEEauthorblockA{\textsuperscript{1}\textit{Department of Computer Science}, Texas A\&M University, College Station, USA}\\
\IEEEauthorblockA{\textsuperscript{2}NVIDIA Corporation, Santa Clara, USA}\\
\IEEEauthorblockA{\textsuperscript{3}\textit{Computing and Artificial Intelligence Division (CAI-3)}, Los Alamos National Laboratory, Los Alamos, USA}
\thanks{* sfeeney@tamu.edu}
}

\maketitle

\begin{abstract}
In this paper, we present SCALAR (Symbolic Conjecture and LLM-Assisted Reasoning), a neurosymbolic framework for automated conjecture generation in quantum circuit analysis built on top of the CUDA-Q open source framework. The system integrates quantum simulation, symbolic conjecture generation, and LLM-based interpretation. We evaluate SCALAR on 82 MaxCut instances from the MQLib benchmark dataset and extend the analysis to 2,000 randomly generated graphs across four topologies: regular, Erdős–Rényi, Barabási–Albert, and Watts–Strogatz. The framework generates conjectured bounds relating optimal QAOA parameters to graph invariants, including known relationships such as periodicity constraints on the phase separation parameter $\gamma$. SCALAR also recovers previously reported parameter transfer phenomena across structurally similar instances. Additionally, the system identifies correlations between graph structural features and optimization landscape properties, which we characterize through invariant-based descriptors. Using CUDA-Q tensor network simulator, we scale experiments to instances of up to 77 qubits. 
We discuss the accuracy, generality, and limitations of the generated conjectures, including sensitivity to graph class and quantum circuit depth.

\end{abstract}


\section{Introduction}
\label{sec:intro}
Quantum computing and artificial intelligence represent two of the most rapidly advancing frontiers in computational science. Their intersection has emerged as a substantial area of interest for researchers in academia, industry, and government \cite{alexeev2025artificial,genesis2025}. A central challenge in near-term quantum computing is the classical optimization burden imposed by variational quantum algorithms: finding optimal parameters requires many expensive quantum evaluations, and the optimization landscape is often non-convex and prone to barren plateaus \cite{larocca2025barren,anschuetz2022quantum}. Understanding how properties of the problem instance (e.g., interaction structure of the cost Hamiltonian) govern these landscapes, and whether optimal parameters can be inferred without repeated optimization, is therefore of fundamental importance.

Artificial intelligence and machine learning approaches for scientific and mathematical discovery have driven breakthroughs in domains ranging from protein structure prediction \cite{jumper2021highly} to advancing research-level mathematics \cite{davies2021advancing, mishra2023mathematicalconjecturegenerationusing,feng2026semi}. In the quantum computing domain, these methods have been applied to circuit synthesis and compilation, the task of translating a target quantum operation into a sequence of hardware-native gates that can be executed on a real device \cite{yan2025quantumcircuitsynthesiscompilation}. They have also been applied to circuit generation, where rather than starting from a known target operation, the goal is to search over candidate circuit architectures to find one that performs well for a given problem \cite{beaudoin2025qfusiondiffusingquantumcircuits,tyagin2025qaoa}. However, these approaches treat parameter optimization as a black box: they can construct or optimize circuits, but offer little insight into \emph{why} optimal parameters take the values they do, or how problem structure governs the optimization landscape. More broadly, the field lacks tools for automated, scalable reasoning about quantum circuit structure and behavior. This work is a first step toward filling this gap.
 
In this paper we introduce \textbf{SCALAR} (Symbolic Conjecture and LLM-Assisted Reasoning Loop), an iterative neurosymbolic framework for quantum circuit analysis. We use the term neurosymbolic to denote the integration of symbolic conjecture generation with neural reasoning components in an iterative discovery process, where each informs subsequent experimental design. We demonstrate its application to the Quantum Approximate Optimization Algorithm (QAOA) \cite{farhi2014quantum} on the MaxCut optimization problem. Our framework proceeds iteratively: quantum circuit simulations produce structured data for a knowledge base containing problem instance invariants (i.e., $n$ the number of nodes), and circuit simulation results; automated conjecture generation (via the txGraffiti system \cite{davila2024automatedconjecturingmathematicsemphtxgraffiti})
conjectures both linear and non-linear bounds over this table. An LLM layer then interprets and ranks conjectures by tightness. Those conjectures are then used to inform the next experimental design cycle, (e.g., the optimal parameter $\gamma^*$ initialized as a function of another optimal parameter $\beta^*$). A key methodological insight of our work is that violation patterns, rows in the knowledge table that violate the bound, in failed conjectures carry scientific signal as well. The central empirical finding of this paper, that optimal QAOA parameters $(\gamma^*, \beta^*)$ are predictable from a small set of graph-structural invariants for graphs sharing a structural fingerprint, emerged directly from analyzing which instances violated a conjectured bound on $\gamma^*$, revealing that all violating instances shared
the same values for their graph invariants
and, upon inspection, identical optimal parameters.

The contributions of this paper are as follows:
 
\begin{enumerate}
\item We introduce \textbf{SCALAR}, a novel, iterative neurosymbolic
        framework for quantum circuit analysis, combining quantum simulation,
        automated symbolic conjecture generation with TxGraffiti, and LLM-assisted interpretation.
        A key design principle is that instances violating a conjecture are
        treated as experimental signals rather than failures, and are used to
        guide the next iteration of the discovery loop.
  \item We apply SCALAR to QAOA for MaxCut on 82 MQLib \cite{dunning2018works} benchmark instances,
        producing a family of analytical bounds relating optimal parameters
        $(\gamma^*, \beta^*)$ to graph-theoretic invariants, including a novel
        lower bound on $\gamma^*$ from chromatic number alone, and rediscovering
        the periodicity of QAOA angles as an emergent conjecture.
\item We present QAOA parameter predictability: optimal parameters $(\gamma^*,
        \beta^*)$ are determined by four cheap graph invariants $(n,
        \bar{d}, \bar{c}, \alpha_{mis})$ across graphs sharing a structural fingerprint
        at circuit depths $p \leq 2$ (where $p$ denotes the number of QAOA rounds,
        formally defined in Section~\ref{sec:bg_qaoa}), a finding that emerged
        directly from analyzing conjecture violations.
\item We apply SCALAR to 1000 randomly generated graphs across
        four topology models (Barab\'{a}si-Albert, Watts-Strogatz, GNM, and
        regular), extending the parameter predictability finding to diverse
        graph topologies and revealing topology-specific conjecture structure,
        including an $O(n^2)$ empirical lower bound on optimizer calls and a
        sign reversal in the degree-$\gamma^*$ relationship unique to small-world
        graphs.


\end{enumerate}
 
The remainder of this paper is organized as follows. Section~\ref{sec:related}
reviews related work. Section~\ref{sec:background} provides background on QAOA,
MaxCut, and txGraffiti. Section~\ref{sec:methodology} describes the framework and
experimental setup. Section~\ref{sec:results} presents our conjecture and
universality results. Section~\ref{sec:discussion} discusses implications, limitations,
and future directions.

The SCALAR framework and experimental code are publicly available at: \href{https://github.com/sfeeney1897/SCALAR}{https://github.com/sfeeney1897/SCALAR}
\section{Related Work}
\label{sec:related}

\subsection{QAOA Parameter Concentration and Transferability}

A growing body of work has studied the phenomenon of QAOA parameter concentration,
wherein optimal parameters for large random instances of a problem class concentrate
around a common value \cite{zhou2020quantum, brandao2018fixed}. Parameter transferability
studies have extended this, showing that parameters optimized on small or random
instances can be transferred to related instances with limited performance loss,
with transferability explained by local subgraph structure and graph parity \cite{galda2023similarity}. Our work is complementary: rather than studying concentration in the 
probabilistic or thermodynamic limit, we study the structural conditions, 
expressible as four graph invariants computable in polynomial time, under 
which graphs sharing a common fingerprint, exhibit matching optimal parameters 
at $p = 1$ across graphs that are not necessarily isomorphic, with evidence 
that the required invariant set grows with circuit depth.
 
\subsection{Automated Mathematical Conjecture Generation}
 
The Graffiti program of Fajtlowicz \cite{FAJTLOWICZ1988113} and its successors,
including Graffiti.pc \cite{DeLaVina2005GraffitiPC} and txGraffiti
\cite{davila2024automatedconjecturingmathematicsemphtxgraffiti}, established the
paradigm of data-driven symbolic conjecture generation in graph theory. These systems
filter a database of objects and their invariants (a knowledge table) through
heuristics such as Dalmatian filtering to produce non-trivial symbolic inequalities.
The GraphCalc package \cite{Davila2025} extends this infrastructure with efficient
computation of graph invariants. More broadly, machine intelligence for mathematical
conjecture generation has been studied in \cite{mishra2023mathematicalconjecturegenerationusing},
and AI-guided mathematical discovery at research level has been demonstrated by
Davies et al.\ \cite{davies2021advancing} and Feng et al. \cite{feng2026semi}. Our work applies this conjecture-generation
paradigm to a new domain: the knowledge table of quantum circuit behavior.
 
\subsection{Quantum Architecture Search and Circuit Discovery}
 
Quantum architecture search (QAS) aims to automate the design of quantum circuit
ans\"{a}tze \cite{10821367}. Related work on circuit synthesis and compilation
\cite{yan2025quantumcircuitsynthesiscompilation} and generative approaches such as
Q-Fusion \cite{beaudoin2025qfusiondiffusingquantumcircuits} and QAOA-GPT and the Generative Quantum Eigensolver \cite{tyagin2025qaoa,nakaji2024generative} similarly seek to reduce
the manual effort of quantum circuit design. AlphaEvolve \cite{novikov2025alphaevolve}
demonstrates the broader power of AI-driven discovery in algorithmic and scientific
settings. In contrast to prior data-driven approaches in quantum architecture search—such as circuit synthesis and generative design, which focus on constructing circuit architectures, our framework targets automated reasoning about quantum circuit behavior. Through an iterative loop of symbolic conjecture generation and LLM-based interpretation, it explores relationships between problem instance structure and the resulting optimization landscape, rather than searching over circuit designs.

\section{Background}
\label{sec:background}
 
\subsection{Classical MaxCut}

The MaxCut problem is a fundamental combinatorial optimization problem.
Given an unweighted graph $G = (V, E)$, the goal is to partition the
vertex set $V$ into two disjoint subsets $S$ and $\bar{S}$ such that
the number of edges crossing the partition is maximized. For a graph
with $n=|V|$ vertices, a feasible solution is encoded as a bit string
$b \in \{0,1\}^n$, where the $j$-th bit indicates to which subset the vertex
$j$ belongs to.

Formally, the MaxCut problem can be expressed as:
\begin{equation}
    \text{MaxCut}(G) = \max_{b \in \{0,1\}^n} \text{cut}(b),
\end{equation}
where
\begin{equation}
    \text{cut}(b) =  \sum_{(i,j) \in E} \mathbf{1}[b_i \neq b_j].
\end{equation}
Here, $\mathbf{1}[b_i \neq b_j]$ is the indicator function that
evaluates to $1$ if vertices $i$ and $j$ are assigned to different
subsets, and $0$ otherwise. In this work we consider only unweighted
graphs, so all edge weights are unity. The exact MaxCut value is
computed by brute-force enumeration over all $2^n$ bitstrings, which
is feasible for the problem sizes considered here ($n \leq 20$).
 
\subsection{QAOA for MaxCut}
\label{sec:bg_qaoa}

The Quantum Approximate Optimization Algorithm (QAOA) \cite{farhi2014quantum} is a
variational quantum-classical hybrid algorithm designed for combinatorial optimization.
QAOA is defined by a cost Hamiltonian $H_C$ encoding the classical objective, a mixing
Hamiltonian $H_M$, a circuit depth $p$, parameter vectors $\boldsymbol{\gamma} =
(\gamma_1, \ldots, \gamma_p)$ and $\boldsymbol{\beta} = (\beta_1, \ldots, \beta_p)$,
and an initial state $|s_0\rangle$ taken to be the uniform superposition
$|{+}\rangle^{\otimes n}$.

For MaxCut on a graph $G = (V, E)$, the cost Hamiltonian is
\begin{equation}
  H_C = \sum_{(u,v)\in E} \frac{1}{2}(1 - Z_u Z_v),
\end{equation}
and the mixer is the standard transverse-field Hamiltonian
\begin{equation}
  H_M = \sum_{v \in V} X_v,
\end{equation}
where $X_v$ and $Z_v$ denote the Pauli-$X$ and Pauli-$Z$ operators on qubit $v$.
This mixer generates transitions between computational basis states and preserves
exploration of the full Hilbert space. More general mixers may be employed provided
they do not commute with $H_C$, though in this work we use the standard transverse-field
formulation.

The QAOA circuit of depth $p$ prepares the state
\begin{equation}
  |\boldsymbol{\gamma}, \boldsymbol{\beta}\rangle
  = e^{-i\beta_p H_M} e^{-i\gamma_p H_C}
    \cdots
    e^{-i\beta_1 H_M} e^{-i\gamma_1 H_C} |{+}\rangle^{\otimes n},
\end{equation}
with parameters\footnote{See Section 7 of the Supplementary Materials of \cite{TATE2026115571} for a proof that these parameter ranges are sufficient.} 
$\boldsymbol{\gamma} \in [0, \pi]^p$ and $\boldsymbol{\beta} \in [0, \pi/2]^p$ optimized classically to maximize the expected cut value
\[
\langle H_C \rangle =
\langle \boldsymbol{\gamma},\boldsymbol{\beta}| H_C |\boldsymbol{\gamma}, \boldsymbol{\beta}\rangle.
\]
In this work, parameters are optimized using the Nelder-Mead method, initialized uniformly at random within the admissible ranges, with these bounds enforced throughout optimization. No post-processing or angle normalization (e.g., periodic wrapping) is applied; all reported parameters correspond directly to the optimizer output.
The approximation ratio is defined as
\begin{equation}
  r = \frac{\langle H_C \rangle_{\text{opt}}}{\text{MaxCut}(G)}.
\end{equation}
In this work, we use $(\gamma^*, \beta^*)$ to denote the numerically optimized parameters found via
Nelder-Mead \cite{10.1093/comjnl/7.4.308}, a gradient-free method, applied to the variational objective, with simulations
carried out using CUDA-Q \cite{The_CUDA-Q_development_team_CUDA-Q}. Due to the complexity of the optimization landscape, we note that Nelder-Mead may not always find the globally optimal values of $\gamma$ and $\beta$.

\subsection{txGraffiti: Automated Conjecture Generation}
\label{sec:bg_txgraffiti}
 
TxGraffiti is a heuristics-based, data-driven Python library for automated mathematical
conjecture generation \cite{davila2024automatedconjecturingmathematicsemphtxgraffiti}.
The name pays homage to a lineage of conjecture generation programs rooted in graph
theory, originating with Fajtlowicz's Graffiti \cite{FAJTLOWICZ1988113} and continuing
with Graffiti.pc \cite{DeLaVina2005GraffitiPC}, both developed at the University of Houston.
 
The core input to txGraffiti is a \emph{knowledge table}: a structured dataset in which
rows represent mathematical objects and columns represent their properties, typically
invariants or any computable feature of interest to the researcher. The core output is
a set of \emph{symbolic conjecture expressions}: inequalities of the form
$f(\text{invariants}) \leq \text{target}$ or $f(\text{invariants}) \geq \text{target}$
that hold (subject to configurable violation tolerance) across all objects in the table.
 
A central challenge in conjecture generation is filtering the combinatorially large
space of candidate inequalities. TxGraffiti addresses this through Dalmatian filtering
and the Morgan filter \cite{davila2024automatedconjecturingmathematicsemphtxgraffiti},
which suppress trivially true or uninteresting conjectures and enforce that each
conjecture must be tight on a minimum number of instances (\texttt{min\_touches}). Two
conjecture modes are used in this work: Graffiti3 (polynomial and composite
functional forms) and ConjecturePlayground (convex hull and ratio methods).

In our application, each row of the knowledge table is a
quantum circuit instance (a graph G with its QAOA simulation results), and the columns contain both static features (graph instance-theoretic invariants) and dynamic features (circuit output quantities).
This is described in detail in Section~\ref{sec:methodology}.
 
\subsection{Graph-Theoretic Invariants}
\label{sec:bg_invariants}

The following graph invariants are used as features in our knowledge table and in the
conjectures discovered by the framework. These invariants were selected to provide a
compact, interpretable set of structural descriptors capturing complementary aspects
of graph topology, including density, local structure, coloring complexity, and
independence structure, while remaining computationally tractable across all instances.

For a graph $G = (V, E)$ with $n = |V|$ vertices and $m = |E|$ edges:

\begin{itemize}
  \item \textbf{Mean degree} $\bar{d} = \frac{2m}{n}$: average vertex degree, capturing overall graph density.
  \item \textbf{Mean clustering coefficient} $\bar{c}$: average local clustering over all vertices, measuring the prevalence of triangle structure \cite{watts1998collective}.
  \item \textbf{Chromatic number} $\chi(G)$: minimum number of colors needed for a proper vertex coloring, reflecting global constraint complexity.
  \item \textbf{Maximum independent set ratio} $\alpha_{\text{mis}} = \alpha(G)/n$: normalized size of the maximum independent set, capturing sparsity and complement structure.
  \item \textbf{Degree assortativity} $r_{\text{assort}}$: Pearson correlation of degrees across edges, measuring degree–degree correlations \cite{newman2003assortativity}.
\end{itemize}

These invariants form the core feature set used in the initial knowledge table construction.
While additional invariants are available (e.g., through MQLib), we restrict to a compact subset
to balance interpretability and computational cost, and to avoid overfitting conjectures to a
large feature space.

For smaller instances ($n \leq 20$), invariants such as the maximum independent set are computed
approximately using NetworkX \cite{hagberg2007exploring}. For larger-scale experiments,
including the MQLib \cite{dunning2018works} benchmark instances and randomly generated graph datasets,
we incorporate precomputed invariants when available and otherwise restrict to efficiently
computable features.

We note that not all invariants are included in the structural fingerprint used for parameter
predictability. In particular, chromatic number $\chi(G)$ is excluded from the initial fingerprint
due to its higher computational cost and limited additional discriminative power relative to the
selected invariants in the low-depth regime; however, it is retained in the knowledge table for
conjecture generation and analysis.


\section{Methodology}
\label{sec:methodology}
 

\subsection{Framework Overview}
 
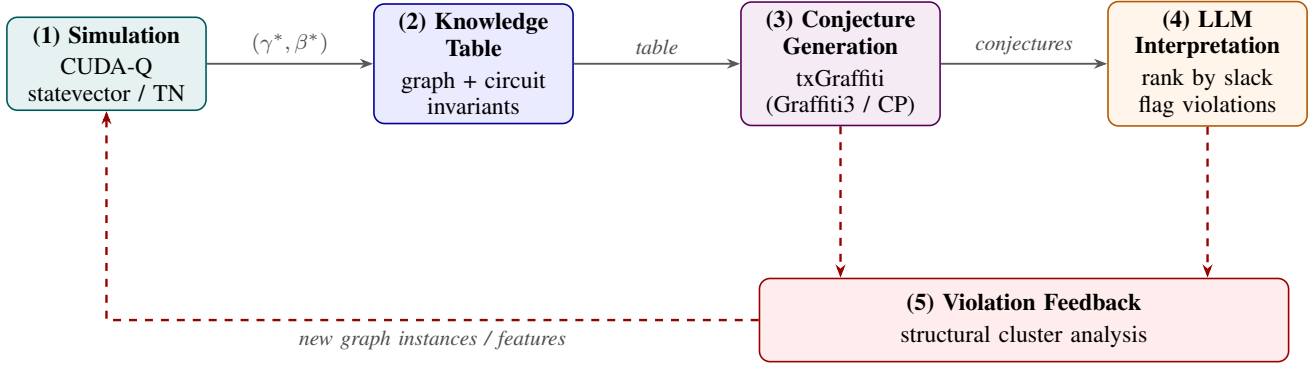
\begin{figure*}[t]
\centering
\begin{tikzpicture}[
  node distance = 1.6cm and 2.2cm,
  box/.style = {
    rectangle, rounded corners=4pt,
    minimum width=2.6cm, minimum height=1.1cm,
    text centered, text width=2.4cm,
    font=\small, line width=0.6pt
  },
  wbox/.style = {
    rectangle, rounded corners=4pt,
    minimum width=2.6cm, minimum height=1.1cm,
    text centered, text width=4cm,
    font=\small, line width=0.6pt
  },
  simbox/.style    = {box, draw=teal!70!black,  fill=teal!10},
  ktbox/.style     = {box, draw=blue!60!black,  fill=blue!8},
  txbox/.style     = {box, draw=violet!70!black,fill=violet!8},
  llmbox/.style    = {box, draw=orange!70!black,fill=orange!8},
  fbbox/.style     = {wbox, draw=red!60!black,   fill=red!7,
                      minimum width=7cm, minimum height=1.1cm},
  arrow/.style     = {-{Stealth[length=5pt]}, line width=0.7pt, gray!70!black},
  feedarrow/.style = {-{Stealth[length=5pt]}, line width=0.9pt,
                      dashed, red!60!black},
  lbl/.style       = {font=\footnotesize\itshape, text=gray!60!black},
]
\node[simbox] (sim)
  {\textbf{(1) Simulation}\\[2pt]
   CUDA-Q\\
   statevector / TN};
\node[ktbox,  right=of sim] (kt)
  {\textbf{(2) Knowledge}\\
   \textbf{Table}\\[2pt]
   graph + circuit\\
   invariants};
\node[txbox,  right=of kt]  (tx)
  {\textbf{(3) Conjecture}\\
   \textbf{Generation}\\[2pt]
   txGraffiti\\
   (Graffiti3 / CP)};
\node[llmbox, right=of tx]  (llm)
  {\textbf{(4) LLM}\\
   \textbf{Interpretation}\\[2pt]
   rank by slack\\
   flag violations};
 
\draw[arrow] (sim) -- node[above,lbl] {$(\gamma^*,\beta^*)$} (kt);
\draw[arrow] (kt)  -- node[above,lbl] {table}               (tx);
\draw[arrow] (tx)  -- node[above,lbl] {conjectures}         (llm);
 
\node[fbbox, below=2.0cm of $(tx.south)!0.5!(llm.south)$] (fb)
  {\textbf{(5) Violation Feedback}\\[2pt]
   structural cluster analysis};
 
\draw[feedarrow] (tx.south)  -- (tx.south  |- fb.north);
\draw[feedarrow] (llm.south) -- (llm.south |- fb.north);
 
\coordinate (bot) at ($(fb.west) + (0, 0)$);
\draw[feedarrow]
  (fb.west) -- (bot)
  -- node[below, lbl, pos=0.5] {new graph instances / features}
     (bot -| sim.south)
  -- (sim.south);
\end{tikzpicture}
\caption{The proposed human-in-the-loop AI-symbolic reasoning framework.
Quantum circuit simulations (Step~1) populate a knowledge table of graph-theoretic
and circuit invariants (Step~2), which is processed by txGraffiti to generate
symbolic conjecture expressions (Step~3). An LLM interprets and ranks conjectures
by slack and tightness, flagging violation patterns (Step~4). Crucially,
violation patterns are analyzed structurally and fed back into the next
experimental design cycle (Step~5, dashed). The parameter predictability finding
reported in Section~\ref{sec:predictability} emerged directly from this feedback loop.}
\label{fig:framework}
\end{figure*}

Our framework is an iterative pipeline consisting of five stages, illustrated in
Fig.~\ref{fig:framework}.
 
\begin{enumerate}
  \item \textbf{Simulation.} Quantum circuit simulations are executed using CUDA-Q
        \cite{The_CUDA-Q_development_team_CUDA-Q} with either a statevector backend
        (exact expectation values, feasible at $n \leq 20$) or a tensor network backend
        (approximate but scalable to $n \lesssim 100$), producing numerically  optimized parameters $(\gamma^*, \beta^*)$
        for each instance.
  \item \textbf{Knowledge table construction.} Graph-theoretic (static features) and circuit invariants (dynamic features)
        are assembled into a structured knowledge table, one row per problem instance.
 \item \textbf{Conjecture generation.} TxGraffiti (Graffiti3 and ConjecturePlayground
modes) is run over the knowledge table, generating both linear and non-linear symbolic inequality conjectures relating invariants to target quantities (e.g., $\gamma^*$, $|\beta^*|$\footnote{We report $|\beta^*|$ due to symmetry in the QAOA landscape: for the MaxCut Hamiltonian with the standard mixer, the objective is invariant under $\beta \mapsto -\beta$, making the sign of $\beta^*$ non-identifying for the relationships studied.}), subject to configurable violation tolerance. In Phase 1, TxGraffiti generates on the order of $10^3$ candidate conjectures, which are filtered down to a smaller set (tens to low hundreds) based on violation tolerance and slack criteria.

\item \textbf{Reasoning and interpretation.} An LLM-based reasoning layer interprets the
conjecture output, ranks conjectures by slack and tightness, highlights informative
bounds, and flags violation patterns for further investigation. In this work we
instantiate this layer using Claude Sonnet 4.6 (Anthropic), though the framework is agnostic
to the choice of reasoning system. Formal methods, human expert review, or other automated reasoners
could serve equally well. The LLM-based layer further prioritizes a subset of conjectures as informative based on tightness and structural interpretability.

\item \textbf{Violation feedback.} Instances that violate or nearly violate conjectures are analyzed structurally. These violation patterns are treated as informative signals rather than failures, guiding the design of follow-on experiments, including the selection of new graph instances and the introduction of additional invariants.
\end{enumerate}

A key design principle of the framework is that \emph{violations are not failures}. 
When a conjecture nearly holds but is violated by a consistent set of instances, these violations may indicate that the current feature set is incomplete or that the dataset does not sufficiently capture the relevant structural variation. 
Analyzing such cases guides both the introduction of additional invariants and the design of new experiments, including expanding the set of graph instances. 
In our setting, this process led directly to the identification of parameter predictability, as described in Section~\ref{sec:predictability}.
 
\subsection{Dataset}

Our experiments are conducted in two phases, reflecting the iterative nature of
the SCALAR framework.

\textbf{Phase 1: MQLib benchmark instances.} We construct the Phase 1 dataset by filtering instances from the MQLib benchmark suite. First, we select all instances with $n \leq 20$ vertices, so that exact MaxCut values can be computed by brute-force enumeration and QAOA can be simulated using a statevector backend on a standard laptop. From this subset, we further restrict to instances with unit edge weights to simplify the MaxCut Hamiltonian. This filtering procedure yields a total of 82 graph instances.

For each instance $G$, we run QAOA at depths $p = 1$ and $p = 2$ using CUDA-Q with a statevector backend. Parameters $(\gamma^*, \beta^*)$ are obtained via Nelder-Mead optimization. We report $|\beta^*|$ throughout for consistency, as the optimization is performed over the restricted domain $\beta \in [0, \pi/2]$, and the sign of $\beta^*$ is therefore not explored in our parameterization.

\textbf{Phase 2: Scale-up to diverse graph models.} To validate conjecture
generalizability beyond the MQLib benchmark, we use NetworkX to generate 1000 random graphs
across four topology models: Barab\'{a}si–Albert (scale-free), Watts–Strogatz
(small-world), GNM/Erd\H{o}s–R\'{e}nyi (structureless random), and $d$-regular
(maximally uniform), with $n$ ranging from 6 to 20. For each model, standard parameterizations
are used (e.g., fixed attachment number for Barab\'{a}si–Albert, rewiring probability for Watts–Strogatz,
edge probability for Erd\H{o}s–R\'{e}nyi, and fixed degree for regular graphs), with parameters chosen to produce
connected graphs; any disconnected instances are discarded. For Phase 2 instances, QAOA is run to depth $p = 5$ using the
Nelder–Mead optimizer.

\subsection{Knowledge Table Construction}
\label{sec:knowledge_table}

The knowledge table is the central data structure of the SCALAR framework. Each row corresponds to a problem instance, and columns represent graph-theoretic invariants (static features) together with circuit-derived quantities (dynamic features).

The size and composition of the knowledge table evolve across experimental phases. In Phase 1, the table contains 82 rows corresponding to the filtered MQLib benchmark instances. In Phase 2, it is expanded to include 2000 randomly generated graphs across multiple topology classes.

The core feature set used in the initial table consists of a small number of interpretable graph invariants, including $n$, $m$, mean degree $\bar{d}$, mean clustering coefficient $\bar{c}$, chromatic number $\chi$, maximum independent set ratio $\alpha_{\text{mis}}$, and degree assortativity $r_\text{assort}$, along with circuit-derived quantities such as approximation ratio $r$ and optimized QAOA parameters $(\gamma^*, |\beta^*|)$.

Although the MQLib dataset provides a large number of graph invariants, we restrict to a compact subset to balance interpretability and computational cost. While some invariants such as chromatic number and maximum independent set are NP-hard in general, we compute them approximately or leverage precomputed values for small instances. We also limit the feature set to avoid including highly correlated invariants, as many graph invariants capture overlapping structural properties, which can introduce redundancy and reduce the interpretability of generated conjectures. This improves conjecture quality by focusing on a smaller set of complementary features rather than a large, redundant feature space.

Importantly, the feature set is not fixed a priori. As part of the SCALAR feedback loop, violation patterns identified during conjecture analysis motivate the introduction of additional invariants beyond the initial feature set. For example, degree standard deviation $\sigma_d$ is introduced in Phase 2 to resolve structural ambiguities between graph classes that share identical values of the initial invariants.

\subsection{Conjecture Generation Configuration}
 
TxGraffiti was run with the following configuration choices, refined through
iterative experimentation:
 
\begin{itemize}
  \item \texttt{max\_violations=2-4}: A tolerance of zero violations is too strict
        for quantum simulation data, which contains small numerical errors from
        grid-search discretization. Allowing 2--4 violations produces tight,
        near-universal conjectures while tolerating numerical noise.
  \item \texttt{min\_touches=1}: Required to ensure all 82 instances participate
        in conjecture generation; higher values suppressed valid conjectures.
  \item Graffiti3 mode: Used for polynomial and composite functional forms, targeting
        $\gamma^*$ as the output quantity with features including $|\beta^*|$,
        $\chi$, $n$, $m$, and $\bar{d}$.
  \item ConjecturePlayground mode: Used with convex hull and ratio methods, targeting
        $\gamma^*$ with features $|\beta^*|$, $\bar{d}$, $m$, and $n$.
\end{itemize}
 To quantify how well conjectures are satisfied, we use the notion of slack. 
For a conjecture of the form $f(x) \leq y$, slack is defined as $y - f(x)$; 
for $f(x) \geq y$, slack is defined as $f(x) - y$. Negative slack indicates a violation.
\section{Results}
\label{sec:results}

\subsection{Analytical Bounds from Conjecture Generation}
\label{sec:conjectures}

Running the SCALAR pipeline on the 82-instance Phase 1 knowledge table yields a set of symbolic conjectures relating $\gamma^*$ to graph-theoretic and circuit-derived quantities. Table~\ref{tab:conjectures} summarizes the primary conjectures, together with their slack statistics. We summarize several observed properties of these conjectures.

\begin{table*}[t]
\centering
\caption{Conjectures generated on the 82-instance Phase 1 dataset.
Mean slack and minimum slack are reported across all instances. Negative minimum
slack indicates small violations (permitted by \texttt{max\_violations} $>0$).
C4$^\dagger$ is included as a sanity check.}
\label{tab:conjectures}
\begin{tabular}{llccc}
\toprule
\textbf{ID} & \textbf{Conjecture} & \textbf{Type} & \textbf{Mean slack} & \textbf{Min slack} \\
\midrule
C1 & $\gamma^* \geq \frac{1}{3}|\beta^*|^2 - \frac{14}{15}|\beta^*| + 2$ & Lower (quadratic) & 0.89 & $-0.008$ \\
C2 & $\gamma^* \geq \frac{45}{13}\chi^2 - 4\chi + \frac{5}{2}$ & Lower (chromatic) & 0.93 & $-0.012$ \\
C3 & $\gamma^* \leq -\frac{1}{18}|\beta^*|^2 + \frac{1}{4}|\beta^*| + \frac{23}{8}$ & Upper (quadratic) & 0.49 & $-0.026$ \\
C4$^\dagger$ & $\gamma^* \leq \frac{22}{7} \approx \pi$ & Upper (trivial) & --- & $0$ \\
C5 & $\gamma^* \geq -2.484\bar{d} + 3.287 - 0.015m$ & Lower (degree/edges) & 0.791 & $0$ \\
C6 & $\gamma^* \geq -2.270\bar{d} + 3.164 - 0.014m$ & Lower (degree/edges) & 0.784 & $0$ \\
C7 & $\gamma^* \leq 0.155|\beta^*| + 2.904$ & Upper (linear) & 0.51 & $\approx 0$ \\
\bottomrule
\end{tabular}
\end{table*}

\textbf{Linear upper bound (C7).} The bound
$\gamma^* \leq 0.155|\beta^*| + 2.904$ holds across all 82 instances with near-zero minimum slack. The expression depends only on $|\beta^*|$ and does not include graph-structural quantities such as $n$ or $m$.

\textbf{Chromatic-number bound (C2).} The inequality
$\gamma^* \geq \frac{45}{13}\chi^2 - 4\chi + \frac{5}{2}$ involves only the chromatic number $\chi$ and no circuit-derived quantities. This indicates that, within this dataset, chromatic number alone provides a lower bound on $\gamma^*$.

\textbf{Trivial constraint (C4).} The bound $\gamma^* \leq 22/7 \approx \pi$ reflects the parameter range imposed during optimization. For unweighted maxcut instances this $\gamma$ is known to be $\gamma \in [0, \pi]$ \cite{sack2021quantum}. Its appearance serves as a consistency check on the conjecture generation process.

\textbf{Degree-dependent bounds (C5--C6).} These inequalities include negative coefficients on $\bar{d}$, indicating that larger mean degree is associated with smaller lower bounds on $\gamma^*$ within this dataset.

\subsection{QAOA Parameter Predictability}
\label{sec:predictability}

This analysis is motivated by investigating violations of conjecture C1. 
Slack analysis showed that all violating instances shared the same values of 
$(n, \bar{d}, \bar{c}, \alpha_{\text{mis}}) = (17, 0.5, 0.429, 0.176)$. 
These graphs are topologically distinct (i.e., not guaranteed to be isomorphic) and exhibit indistinguishable optimized parameters $(\gamma^*, \beta^*)$ at $p = 1$. In some cases, graphs within a fingerprint group are isomorphic, for which parameter agreement is expected; however, we also observe agreement across instances that are not trivially identical under isomorphism.

This observation motivates a systematic analysis. We partition the 82 instances 
by the structural fingerprint $(n, \bar{d}, \bar{c}, \alpha_{\text{mis}})$ and 
compute the standard deviation of $\gamma^*$ and $\beta^*$ within each group 
of size $\geq 2$. The results are shown in Table~\ref{tab:universality}.

Across the 14 repeated-fingerprint groups, all but one group have near-zero variance 
($\sigma \approx 0$) in both $\gamma^*$ and $\beta^*$ at $p = 1$ and $p = 2$. 
These groups span problem sizes $n \in \{6, 8, 10, 11, 12, 13, 17, 18, 20\}$; the one exceptional group has problem size $n=14$.

We summarize this observation as an empirical pattern:

\begin{quote}
\textbf{Empirical observation (Parameter predictability).}
For $p = 1$ and $p = 2$ QAOA on unweighted MaxCut, graphs sharing the same 
values of the structural fingerprint $(n, \bar{d}, \bar{c}, \alpha_{\text{mis}})$ 
often exhibit numerically similar optimized parameters $(\gamma^*, \beta^*)$, 
up to optimization tolerance and potential landscape degeneracy.
\end{quote}

This suggests that, within the Phase 1 dataset, optimal parameters are largely 
determined by a small set of graph invariants. One possible implication is that 
parameters may be reused across instances with matching fingerprints, reducing 
the need for repeated optimization. However, this behavior is not uniform across 
all instances and is examined further in the Phase 2 analysis.

A limitation of this result is that many fingerprint groups in the Phase 1 dataset 
contain isomorphic graphs, for which parameter agreement is expected \cite{shaydulin2021classical}. Only two 
groups ($n = 13$ and $n = 18$) are confirmed to contain non-isomorphic graphs 
with matching optimized parameters.

The previously mentioned $n = 14$ exception is discussed in 
Section~\ref{sec:exception}, and the generality of this pattern is evaluated 
in Section~\ref{sec:invariants}.

\begin{table*}[t]
\centering
\caption{Standard deviation of optimal QAOA parameters within each structural
fingerprint group $(n, \bar{d}, \bar{c}, \alpha_{\text{mis}})$ having two or more
instances (14 groups total). All $\sigma$ values are computed across topologically
distinct graphs sharing the same fingerprint. Thirteen of 14 groups show
$\sigma \approx 0$ at both $p=1$ and $p=2$, confirming parameter universality.
The single exception ($n=14$, $\bar{d}=0.692$, highlighted) exhibits two distinct
optimization landscape basins and is discussed in Section~\ref{sec:exception}.}
\label{tab:universality}
\setlength{\tabcolsep}{5pt}
\begin{tabular}{cccccrcccc}
\toprule
$n$ & $\bar{d}$ & $\bar{c}$ & $\alpha_{\text{mis}}$ & count
    & $\sigma(\gamma_1^*)$ & $\sigma(\beta_1^*)$
    & $\sigma(\gamma_2^*)$ & $\sigma(\beta_2^*)$ \\
\midrule
 6 & 0.600 & 0.000 & 0.500 & 3 & $9.7 \times 10^{-5}$ & $6.7 \times 10^{-5}$ & $1.6 \times 10^{-4}$ & $2.8 \times 10^{-4}$ \\
 8 & 0.429 & 0.000 & 0.500 & 3 & $8.8 \times 10^{-6}$ & $3.5 \times 10^{-5}$ & $1.6 \times 10^{-4}$ & $1.5 \times 10^{-4}$ \\
 8 & 0.750 & 0.600 & 0.375 & 2 & $4.4 \times 10^{-4}$ & $3.6 \times 10^{-5}$ & $4.5 \times 10^{-4}$ & $5.1 \times 10^{-4}$ \\
10 & 0.333 & 0.000 & 0.400 & 5 & $5.0 \times 10^{-4}$ & $2.0 \times 10^{-4}$ & $3.5 \times 10^{-4}$ & $5.6 \times 10^{-4}$ \\
11 & 0.364 & 0.000 & 0.455 & 2 & $0$                  & $0$                  & $0$                  & $0$                  \\
12 & 0.455 & 0.500 & 0.250 & 5 & $9.0 \times 10^{-8}$ & $1.6 \times 10^{-9}$ & $4.5 \times 10^{-8}$ & $2.8 \times 10^{-7}$ \\
13 & 0.500 & 0.452 & 0.231 & 2 & $2.4 \times 10^{-3}$ & $2.0 \times 10^{-3}$ & $7.5 \times 10^{-4}$ & $2.4 \times 10^{-3}$ \\
\rowcolor{red!8}
14 & 0.692 & 0.648 & 0.143 & 2 & $5.2 \times 10^{-1}$ & $1.6 \times 10^{0\phantom{-}}$ & $3.9 \times 10^{-1}$ & $1.3 \times 10^{-1}$ \\
17 & 0.500 & 0.429 & 0.176 & 6 & $5.1 \times 10^{-4}$ & $1.2 \times 10^{-4}$ & $1.0 \times 10^{-3}$ & $5.4 \times 10^{-4}$ \\
18 & 0.314 & 0.533 & 0.278 & 2 & $1.0 \times 10^{-3}$ & $1.1 \times 10^{-3}$ & $1.0 \times 10^{-6}$ & $5.8 \times 10^{-5}$ \\
20 & 0.137 & 0.000 & 0.500 & 2 & $0$                  & $0$                  & $0$                  & $0$                  \\
20 & 0.158 & 0.000 & 0.350 & 2 & $0$                  & $0$                  & $0$                  & $0$                  \\
20 & 0.158 & 0.000 & 0.400 & 3 & $0$                  & $0$                  & $0$                  & $0$                  \\
20 & 0.263 & 0.300 & 0.250 & 4 & $4.1 \times 10^{-4}$ & $6.9 \times 10^{-4}$ & $3.0 \times 10^{-6}$ & $1.0 \times 10^{-4}$ \\
\bottomrule
\end{tabular}
\end{table*}

\subsection{The $n=14$ Exception and Multi-Basin Landscapes}
\label{sec:exception}

A notable exception to the parameter predictability pattern occurs at 
$n = 14$, $\bar{d} = 0.692$. Instances in this fingerprint group do not 
converge to a single consistent set of optimized parameters 
$(\gamma^*, \beta^*)$. Instead, optimization results cluster into two 
distinct regions of parameter space. We verify that these clusters are not related by simple periodic symmetries of the QAOA parameters (e.g., shifts by $\pi$ or sign flips), suggesting that they correspond to distinct optima under the chosen parameterization.

This behavior indicates that the four-invariant fingerprint $(n, \bar{d}, \bar{c}, \alpha_{\text{mis}})$ is insufficient to uniquely determine the optimized parameters for this class of graphs. We emphasize that this observation does not distinguish between multiple possible explanations: the underlying optimization landscape may contain multiple basins, or different instances may exhibit distinct optimal parameters despite sharing the same fingerprint. Our results provide evidence for the latter, namely that the chosen invariant set does not fully capture the structural variation relevant to QAOA parameter selection.

This behavior suggests that the four-invariant fingerprint 
$(n, \bar{d}, \bar{c}, \alpha_{\text{mis}})$ is insufficient to fully 
characterize the optimization landscape for this class of graphs. 
Additional structural features, such as degree assortativity or chromatic 
number, may distinguish between the observed basins, though identifying 
the minimal distinguishing set remains an open question.

\subsection{Phase 2: Generalization and Invariant Refinement}
\label{sec:invariants}

The Phase 1 analysis suggests that QAOA parameters are largely determined by a 
small set of graph invariants. We next test whether this pattern generalizes to a 
broader set of graph structures. To do this, we apply the same four-invariant fingerprint 
$(n, \bar{d}, \bar{c}, \alpha_{\text{mis}})$ to the Phase 2 dataset of 1000 
randomly generated graphs spanning multiple topology models. Repeating the 
groupby analysis shows that the consistency observed in Phase 1 does not fully 
carry over: at $p = 1$, the fraction of repeated-fingerprint groups with 
near-zero parameter variance drops from 92.9\% to 54.0\%.

We examine the source of this discrepancy. In many cases, graphs with identical 
values of $(n, \bar{d}, \bar{c}, \alpha_{\text{mis}})$ produce different optimized 
parameters. A common example occurs between $d$-regular graphs and 
Watts–Strogatz graphs, which can share the same mean degree and clustering 
coefficient but differ in the distribution of degrees.

This observation suggests that the four-invariant fingerprint does not capture 
all relevant structural variation. In particular, degree standard deviation 
$\sigma_d$ distinguishes these cases: it is zero for regular graphs and non-zero 
for Watts–Strogatz graphs with the same mean degree. Incorporating $\sigma_d$ 
as an additional invariant resolves most of these discrepancies. Table~\ref{tab:invariants} summarizes the effect of extending the fingerprint 
with additional invariants across multiple circuit depths.

\begin{table}[t]
\centering
\caption{Universality rates (fraction of repeated-fingerprint groups with 
$\sigma(\gamma^*, \beta^*) < \varepsilon$) for fingerprints of increasing size,
evaluated at circuit depths $p = 1$, $p = 2$, and $p = 5$ on the Phase 2 dataset. 
Here, $\varepsilon$ denotes a small tolerance threshold (set to $10^{-3}$ in our experiments). 
Results are computed within each graph model separately (i.e., instances are grouped by both fingerprint and topology class). 
Adding $\sigma_d$ alone increases $p=1$ universality from 60.9\% to 96.6\%.}
\label{tab:invariants}
\begin{tabular}{lccc}
\toprule
\textbf{Fingerprint} & $p=1$ & $p=2$ & $p=5$ \\
\midrule
4-invariant: $(n, \bar{d}, \bar{c}, \alpha_{\text{mis}})$           & 60.9\% & 16.3\% &  8.7\% \\
5-invariant: $+\,\sigma_d$                                           & 96.6\% & 27.6\% & 11.5\% \\
5-invariant: $+\,r_{\text{assort}}$                                  & ---    & ---    & ---    \\
6-invariant: $+\,\sigma_d +\, r_{\text{assort}}$                    & 98.6\% & 39.4\% & 19.7\% \\
\bottomrule
\end{tabular}
\end{table}

Several observations follow from Table~\ref{tab:invariants}.

\textbf{Degree variance improves separation.} Adding degree standard deviation $\sigma_d$ increases the fraction of repeated-fingerprint groups with near-zero parameter variance from 60.9\% to 96.6\% at $p = 1$, resolving most discrepancies observed with the four-invariant fingerprint.

\textbf{Performance degrades with depth.} Even with additional invariants, this consistency decreases as circuit depth increases, dropping from 98.6\% at $p = 1$ to 39.4\% at $p = 2$ and 19.7\% at $p = 5$. This suggests that higher-depth QAOA parameters depend on more detailed aspects of graph structure. One possible explanation is that increasing circuit depth expands the effective interaction radius of the algorithm, allowing it to capture more global features of the graph. In contrast, many of the invariants used in our feature set (e.g., mean degree, clustering coefficient) primarily encode local or aggregate structural properties. As a result, these invariants may be sufficient to characterize low-depth behavior but become less informative as deeper circuits incorporate longer-range correlations and more complex global structure.

\textbf{First-layer parameters are more stable.} At $p = 5$, restricting attention to $(\gamma_1^*, \beta_1^*)$ yields higher consistency (32.5\%) than considering all parameters jointly (19.7\%).

\textbf{Cross-model agreement at $p = 1$.} For groups containing graphs from different topology models, adding $\sigma_d$ yields consistent parameters at $p = 1$ across all observed cases, whereas the four-invariant fingerprint does not.

\subsection{Additional Patterns at Scale}
\label{sec:global_conjectures}

Applying the framework to the larger Phase 2 dataset also reveals several 
recurring patterns across graph models and sizes. We summarize the most 
consistent observations.

\textbf{Scaling pattern in $\gamma^*$.}
Across multiple conjectures, upper bounds on $\gamma^*$ consistently include 
a leading term proportional to $n$, for example:
\begin{align}
  \gamma^* &\leq \tfrac{1}{4}n + \tfrac{8}{13}\sqrt{n}, \\
  \gamma^* &\leq \tfrac{1}{4}n + \tfrac{7}{6}\sqrt{\bar{d}}.
\end{align}
However, these bounds become vacuous for larger graph sizes. In particular, the right-hand side exceeds the trivial constraint $\gamma^* \leq \pi$ for sufficiently large $n$ (e.g., $n \geq 7$ for the first bound), indicating that these expressions are primarily informative in the small-$n$ regime. This reflects the scale of the instances considered and suggests that the observed scaling behavior is most relevant for small to moderate problem sizes. More generally, it highlights a limitation of unconstrained conjecture generation, which may produce bounds that are locally tight but do not remain informative asymptotically.

\textbf{Approximation ratio bounds.}
The conjectures relate approximation ratio $r$ to structural properties such as 
assortativity and independence number. For example:
\begin{align}
  r &\leq 1 - \tfrac{3}{28}\sqrt{r_{\text{assort}}}, \\
  r &\leq \left\lfloor \tfrac{1}{12}n + \sqrt{\alpha_{\text{mis}}} \right\rfloor.
\end{align}
These expressions indicate that approximation quality varies with graph structure in a measurable way. We note that degree assortativity $r_{\text{assort}}$ may take negative values. The above expression is reported as generated by TxGraffiti from the observed dataset, and is therefore interpreted over the range of values present in that data, not a general prediction.

\textbf{Optimizer call scaling.}
The conjectures include lower bounds on the number of objective function calls 
required by Nelder--Mead optimization:
\begin{equation}
  \tfrac{22}{17}n^2 + 4n + 4 \leq {\text{obj}}.
\end{equation}
This is consistent with the quadratic scaling of simplex-based methods with 
respect to the number of optimization parameters.
\cite{10.1093/comjnl/7.4.308}

\textbf{Topology-dependent behavior.}
When conjectures are evaluated within individual graph models, differences in 
structure become apparent. For example, in Watts--Strogatz graphs, bounds on 
$\gamma^*$ include a negative coefficient on mean degree, whereas other models 
exhibit positive coefficients. This indicates that the relationship between 
graph structure and optimized parameters depends on topology.

\subsection{Scaling Demonstration with Tensor Network Simulation}

To complement the preceding analysis, we perform a proof-of-feasibility experiment
using the CUDA-Q tensor network backend on a single instance with 77 qubits.
The SCALAR pipeline is successfully executed in this setting, producing a valid
objective value and optimized parameters using the same optimization procedure
as in smaller-scale experiments. We emphasize that this is a single-instance
demonstration rather than a systematic scaling study. Nevertheless, it illustrates
that the framework can be instantiated beyond the exact statevector regime.

\section{Discussion}
\label{sec:discussion}
 
\subsection{Main Takeaways}

We summarize the main findings of this work. Conjecture-driven analysis produces interpretable relationships. Applying 
SCALAR to QAOA simulations yields symbolic bounds relating optimal parameters 
to graph structure. These conjectures provide a compact description of patterns 
in the data.

Low-depth QAOA parameters exhibit structured regularity. At $p = 1$ and $p = 2$, 
many graph instances with matching values of a small set of invariants produce 
numerically similar optimized parameters. This suggests that, in this regime, 
QAOA behavior is largely determined by coarse structural features.

Invariant choice is critical and problem-dependent. The Phase 2 analysis shows 
that initial invariants are not sufficient to generalize across graph families. 
Introducing additional features, such as degree variance, improves consistency 
but does not fully resolve the problem, especially at higher circuit depths.

QAOA behavior becomes more instance-specific with depth. As circuit depth 
increases, the consistency of optimized parameters across similar instances 
decreases. This indicates that higher-depth QAOA depends on finer-grained 
structural properties not captured by simple invariant sets.

Automated conjecture generation can guide analysis. The SCALAR workflow 
identifies patterns, highlights exceptions, and suggests directions for 
refinement, providing a structured way to explore relationships between 
quantum circuits and problem structure.

\subsection{Limitations and Caveats}

This work represents an initial step toward automated reasoning about quantum 
circuits, and several limitations should be noted.

All conjectures and patterns reported are empirical and derived from finite 
datasets. They should not be interpreted as general statements about QAOA 
beyond the tested instances and parameter ranges.

The analysis is restricted to unweighted MaxCut at low circuit depths 
($p \leq 2$), where the strongest patterns are observed. At higher depths, 
parameter behavior becomes less consistent, and the current invariant sets 
do not fully capture the underlying structure.

The results depend on the choice of graph invariants used to construct the 
knowledge table. While additional invariants can improve performance, there 
is no guarantee that a small set of features will suffice in general.

Optimized parameters are obtained through numerical methods and may be 
affected by local minima, initialization, and optimizer settings. Observed 
agreement between instances is therefore subject to optimization tolerance 
and potential landscape degeneracy.

Although SCALAR automates conjecture generation, the identification and 
interpretation of meaningful invariants currently requires human input. 
The system is not fully autonomous and relies on domain knowledge to guide 
refinement.

Finally, the datasets considered are limited in scope. Phase 1 uses a small 
set of benchmark instances, and Phase 2 relies on synthetic graph models. 
While these provide useful test cases, they do not cover the full range of 
problem instances encountered in practice.

\subsection{Future Work}
 The conjectures that SCALAR finds with the help of txGraffiti \cite{davila2024automatedconjecturingmathematicsemphtxgraffiti} can be exported to the Lean 4 \cite{deMoura2021lean4} proof assistant.
 Then a formal proof of these conjectures can be formulated and checked in Lean. However, Lean will in general not be able to automatically find proofs of these conjectures, except in very simple cases. One reason is that the grind tactic tries to avoid computationally expensive graph invariants, and other tactics follow a similar approach for efficiency reasons. However, Lean is starting to offer more support for graphs and  graph invariants. Writing custom tactics for automated theorem proving of graphs conjectures is an interesting direction to pursue in the future.

 \section{Code and Data Availability}
 The code, datasets, and experiment configurations used in this work
are available at:
\href{https://github.com/sfeeney1897/SCALAR}{https://github.com/sfeeney1897/SCALAR}

\section*{Acknowledgments}
The research presented in this article was supported by the NNSA’s Advanced
Simulation and Computing Beyond Moore’s Law Program at Los Alamos National Laboratory. LANL report LA-UR-26-23374.
\bibliographystyle{IEEEtran}
\bibliography{sources}{}

\begin{thebibliography}{10}
\providecommand{\url}[1]{#1}
\csname url@samestyle\endcsname
\providecommand{\newblock}{\relax}
\providecommand{\bibinfo}[2]{#2}
\providecommand{\BIBentrySTDinterwordspacing}{\spaceskip=0pt\relax}
\providecommand{\BIBentryALTinterwordstretchfactor}{4}
\providecommand{\BIBentryALTinterwordspacing}{\spaceskip=\fontdimen2\font plus
\BIBentryALTinterwordstretchfactor\fontdimen3\font minus \fontdimen4\font\relax}
\providecommand{\BIBforeignlanguage}[2]{{%
\expandafter\ifx\csname l@#1\endcsname\relax
\typeout{** WARNING: IEEEtran.bst: No hyphenation pattern has been}%
\typeout{** loaded for the language `#1'. Using the pattern for}%
\typeout{** the default language instead.}%
\else
\language=\csname l@#1\endcsname
\fi
#2}}
\providecommand{\BIBdecl}{\relax}
\BIBdecl

\bibitem{alexeev2025artificial}
Y.~Alexeev, M.~H. Farag, T.~L. Patti, M.~E. Wolf, N.~Ares, A.~Aspuru-Guzik, S.~C. Benjamin, Z.~Cai, S.~Cao, C.~Chamberland \emph{et~al.}, ``Artificial intelligence for quantum computing,'' \emph{Nature Communications}, vol.~16, no.~1, p. 10829, 2025.

\bibitem{genesis2025}
{U.S. Department of Energy}, ``{Genesis Mission}: A national mission to accelerate science through artificial intelligence,'' \url{https://genesis.energy.gov/}, 2025, accessed: April 2026.

\bibitem{larocca2025barren}
M.~Larocca, S.~Thanasilp, S.~Wang, K.~Sharma, J.~Biamonte, P.~J. Coles, L.~Cincio, J.~R. McClean, Z.~Holmes, and M.~Cerezo, ``Barren plateaus in variational quantum computing,'' \emph{Nature Reviews Physics}, vol.~7, no.~4, pp. 174--189, 2025.

\bibitem{anschuetz2022quantum}
E.~R. Anschuetz and B.~T. Kiani, ``Quantum variational algorithms are swamped with traps,'' \emph{Nature Communications}, vol.~13, no.~1, p. 7760, 2022.

\bibitem{jumper2021highly}
J.~Jumper, R.~Evans, A.~Pritzel, T.~Green, M.~Figurnov, O.~Ronneberger, K.~Tunyasuvunakool, R.~Bates, A.~{\v{Z}}{\'\i}dek, A.~Potapenko \emph{et~al.}, ``Highly accurate protein structure prediction with alphafold,'' \emph{nature}, vol. 596, no. 7873, pp. 583--589, 2021.

\bibitem{davies2021advancing}
A.~Davies, P.~Veli{\v{c}}kovi{\'c}, L.~Buesing, S.~Blackwell, D.~Zheng, N.~Toma{\v{s}}ev, R.~Tanburn, P.~Battaglia, C.~Blundell, A.~Juh{\'a}sz \emph{et~al.}, ``Advancing mathematics by guiding human intuition with ai,'' \emph{Nature}, vol. 600, no. 7887, pp. 70--74, 2021.

\bibitem{mishra2023mathematicalconjecturegenerationusing}
\BIBentryALTinterwordspacing
C.~Mishra, S.~R. Moulik, and R.~Sarkar, ``Mathematical conjecture generation using machine intelligence,'' 2023. [Online]. Available: \url{https://arxiv.org/abs/2306.07277}
\BIBentrySTDinterwordspacing

\bibitem{feng2026semi}
T.~Feng, T.~Trinh, G.~Bingham, J.~Kang, S.~Zhang, S.-h. Kim, K.~Barreto, C.~Schildkraut, J.~Jung, J.~Seo \emph{et~al.}, ``Semi-autonomous mathematics discovery with gemini: A case study on the erd$\backslash$h $\{$o$\}$ s problems,'' \emph{arXiv preprint arXiv:2601.22401}, 2026.

\bibitem{yan2025quantumcircuitsynthesiscompilation}
\BIBentryALTinterwordspacing
G.~Yan, W.~Wu, Y.~Chen, K.~Pan, X.~Lu, Z.~Zhou, Y.~Wang, R.~Wang, and J.~Yan, ``Quantum circuit synthesis and compilation optimization: Overview and prospects,'' 2025. [Online]. Available: \url{https://arxiv.org/abs/2407.00736}
\BIBentrySTDinterwordspacing

\bibitem{beaudoin2025qfusiondiffusingquantumcircuits}
\BIBentryALTinterwordspacing
C.~Beaudoin and S.~Ghosh, ``Q-fusion: Diffusing quantum circuits,'' 2025. [Online]. Available: \url{https://arxiv.org/abs/2504.20794}
\BIBentrySTDinterwordspacing

\bibitem{tyagin2025qaoa}
I.~Tyagin, M.~H. Farag, K.~Sherbert, K.~Shirali, Y.~Alexeev, and I.~Safro, ``Qaoa-gpt: Efficient generation of adaptive and regular quantum approximate optimization algorithm circuits,'' in \emph{2025 IEEE International Conference on Quantum Computing and Engineering (QCE)}, vol.~1.\hskip 1em plus 0.5em minus 0.4em\relax IEEE, 2025, pp. 1505--1515.

\bibitem{farhi2014quantum}
E.~Farhi, J.~Goldstone, and S.~Gutmann, ``A quantum approximate optimization algorithm,'' \emph{arXiv preprint arXiv:1411.4028}, 2014.

\bibitem{davila2024automatedconjecturingmathematicsemphtxgraffiti}
\BIBentryALTinterwordspacing
R.~Davila, ``Automated conjecturing in mathematics with \emph{TxGraffiti},'' 2024. [Online]. Available: \url{https://arxiv.org/abs/2409.19379}
\BIBentrySTDinterwordspacing

\bibitem{dunning2018works}
I.~Dunning, S.~Gupta, and J.~Silberholz, ``What works best when? a systematic evaluation of heuristics for max-cut and qubo,'' \emph{INFORMS Journal on Computing}, vol.~30, no.~3, pp. 608--624, 2018.

\bibitem{zhou2020quantum}
L.~Zhou, S.-T. Wang, S.~Choi, H.~Pichler, and M.~D. Lukin, ``Quantum approximate optimization algorithm: Performance, mechanism, and implementation on near-term devices,'' \emph{Physical Review X}, vol.~10, no.~2, p. 021067, 2020.

\bibitem{brandao2018fixed}
F.~G. Brandao, M.~Broughton, E.~Farhi, S.~Gutmann, and H.~Neven, ``For fixed control parameters the quantum approximate optimization algorithm's objective function value concentrates for typical instances,'' \emph{arXiv preprint arXiv:1812.04170}, 2018.

\bibitem{galda2023similarity}
A.~Galda, E.~Gupta, J.~Falla, X.~Liu, D.~Lykov, Y.~Alexeev, and I.~Safro, ``Similarity-based parameter transferability in the quantum approximate optimization algorithm,'' \emph{Frontiers in Quantum Science and Technology}, vol.~2, p. 1200975, 2023.

\bibitem{FAJTLOWICZ1988113}
\BIBentryALTinterwordspacing
S.~Fajtlowicz, ``On conjectures of graffiti,'' in \emph{Graph Theory and Applications}, ser. Annals of Discrete Mathematics, J.~Akiyama, Y.~Egawa, and H.~Enomoto, Eds.\hskip 1em plus 0.5em minus 0.4em\relax Elsevier, 1988, vol.~38, pp. 113--118. [Online]. Available: \url{https://www.sciencedirect.com/science/article/pii/S0167506008707763}
\BIBentrySTDinterwordspacing

\bibitem{DeLaVina2005GraffitiPC}
E.~DeLaVi{\~{n}}a, ``Graffiti.pc: A variant of graffiti,'' in \emph{Graphs and Discovery}, ser. DIMACS Series in Discrete Mathematics and Theoretical Computer Science.\hskip 1em plus 0.5em minus 0.4em\relax Providence, RI: American Mathematical Society, 2005, vol.~69, pp. 71--79.

\bibitem{Davila2025}
\BIBentryALTinterwordspacing
R.~Davila, ``Graphcalc: A python package for computing graph invariants in automated conjecturing systems,'' \emph{Journal of Open Source Software}, vol.~10, no. 112, p. 8383, 2025. [Online]. Available: \url{https://doi.org/10.21105/joss.08383}
\BIBentrySTDinterwordspacing

\bibitem{10821367}
D.~Martyniuk, J.~Jung, and A.~Paschke, ``Quantum architecture search: A survey,'' in \emph{2024 IEEE International Conference on Quantum Computing and Engineering (QCE)}, vol.~01, 2024, pp. 1695--1706.

\bibitem{nakaji2024generative}
K.~Nakaji, L.~B. Kristensen, R.~Kemmoku, J.~A. Campos-Gonzalez-Angulo, M.~G. Vakili, H.~Huang, M.~Bagherimehrab, C.~Gorgulla, F.~Wong, A.~McCaskey \emph{et~al.}, ``The generative quantum eigensolver (gqe) and its application for ground state search,'' \emph{arXiv preprint arXiv:2401.09253}, 2024.

\bibitem{novikov2025alphaevolve}
A.~Novikov, N.~V{\~u}, M.~Eisenberger, E.~Dupont, P.-S. Huang, A.~Z. Wagner, S.~Shirobokov, B.~Kozlovskii, F.~J. Ruiz, A.~Mehrabian \emph{et~al.}, ``Alphaevolve: A coding agent for scientific and algorithmic discovery,'' \emph{arXiv preprint arXiv:2506.13131}, 2025.

\bibitem{TATE2026115571}
\BIBentryALTinterwordspacing
R.~Tate and S.~Eidenbenz, ``Theoretical approximation ratios for warm-started qaoa on 3-regular max-cut instances at depth p=1,'' \emph{Theoretical Computer Science}, vol. 1059, p. 115571, 2026. [Online]. Available: \url{https://www.sciencedirect.com/science/article/pii/S0304397525005092}
\BIBentrySTDinterwordspacing

\bibitem{10.1093/comjnl/7.4.308}
\BIBentryALTinterwordspacing
J.~A. Nelder and R.~Mead, ``A simplex method for function minimization,'' \emph{The Computer Journal}, vol.~7, no.~4, pp. 308--313, 01 1965. [Online]. Available: \url{https://doi.org/10.1093/comjnl/7.4.308}
\BIBentrySTDinterwordspacing

\bibitem{The_CUDA-Q_development_team_CUDA-Q}
\BIBentryALTinterwordspacing
{The CUDA-Q development team}, ``{CUDA-Q}.'' [Online]. Available: \url{https://github.com/NVIDIA/cuda-quantum}
\BIBentrySTDinterwordspacing

\bibitem{watts1998collective}
D.~J. Watts and S.~H. Strogatz, ``Collective dynamics of ‘small-world’networks,'' \emph{nature}, vol. 393, no. 6684, pp. 440--442, 1998.

\bibitem{newman2003assortativity}
\BIBentryALTinterwordspacing
M.~E.~J. Newman, ``Mixing patterns in networks,'' \emph{Phys. Rev. E}, vol.~67, p. 026126, Feb 2003. [Online]. Available: \url{https://link.aps.org/doi/10.1103/PhysRevE.67.026126}
\BIBentrySTDinterwordspacing

\bibitem{hagberg2007exploring}
A.~Hagberg, P.~J. Swart, and D.~A. Schult, ``Exploring network structure, dynamics, and function using networkx,'' Los Alamos National Laboratory (LANL), Tech. Rep., 2007.

\bibitem{sack2021quantum}
S.~H. Sack and M.~Serbyn, ``Quantum annealing initialization of the quantum approximate optimization algorithm,'' \emph{quantum}, vol.~5, p. 491, 2021.

\bibitem{shaydulin2021classical}
R.~Shaydulin, S.~Hadfield, T.~Hogg, and I.~Safro, ``Classical symmetries and the quantum approximate optimization algorithm,'' \emph{Quantum Information Processing}, vol.~20, no.~11, p. 359, 2021.

\bibitem{deMoura2021lean4}
L.~de~Moura and S.~Ullrich, ``The {L}ean 4 theorem prover and programming language,'' in \emph{International Conference on Automated Deduction}.\hskip 1em plus 0.5em minus 0.4em\relax Springer, 2021, pp. 625--635.

\end{thebibliography}

\end{document}